\documentclass[12pt,twoside,english]{article}
\usepackage[T1]{fontenc}
\usepackage[latin1]{inputenc}
\usepackage{amsbsy}

\makeatletter

\newcommand{\noun}[1]{\textsc{#1}}

\makeatother

\usepackage{babel}
\begin{document}
\title{The Navier-Stokes equation and a fully developed turbulence}
\author{{\normalsize{}M. Apostol }\\
{\normalsize{}Department of Theoretical Physics, Institute of Atomic
Physics, }\\
{\normalsize{}Magurele-Bucharest MG-6, POBox MG-35, Romania }\\
{\normalsize{}email: apoma@theory.nipne.ro}}
\date{{}}

\maketitle
\relax
\begin{abstract}
In fairly general conditions we give explicit (smooth) solutions for
the potential flow. We show that, rigorously speaking, the equations
of the fluid mechanics have not rotational solutions. However, within
the usual approximations of an incompressible fluid and an isentropic
flow, the remaining Navier-Stokes equation has approximate vorticial
(rotational) solutions, generated by viscosity. In general, the vortices
are unstable, and a discrete distribution of vorticial solutions is
not in mechanical equilibrium; it forms an unstable vorticial liquid.
On the other hand, these solutions may exhibit turbulent, fluctuating
instabilities for large variations of the velocity over short distances.
We represent a fully developed turbulence as a homogeneous, isotropic
and highly-fluctuating distribution of singular centres of turbulence.
A regular mean flow can be included. In these circumstances the Navier-Stokes
equation exhibits three time scales. The equations of the mean flow
can be disentangled from the equations of the fluctuating part, which
is reduced to a vanishing inertial term. This latter equation is not
satisfied after averaging out the temporal fluctuations. However,
for a homogeneous and isotropic distribution of non-singular turbulence
centres the equation for the inertial term is satisfied trivially,
\emph{i.e.} both the average fluctuating velocity and the average
fluctuating inertial term are zero. If the velocity is singular at
the turbulence centres, we are left with a quasi-ideal classical gas
of singularities, or a solution of singularities in quasi thermal
equilibrium in the background fluid. This is an example of an emergent
dynamics. We give three examples of vorticial liquids.
\end{abstract}
\relax


\section{Introduction}

In fairly general conditions we give explicit (smooth) solutions for
the potential flow. As it is well known, the fluids may develop turbulence.
In its extreme manifestation the turbulent flow displays very irregular,
disordered velocities, fluctuating in time at each point in space.
This is known as a fully developed turbulence. By using such fluctuating
velocities, besides a steady mean velocity, the Navier-Stokes equation
becomes an infinite hierarchy of equations for velocity mean correlation
functions, known as Reynolds's equations,\cite{key-1} which need
closure assumptions. According to the experimental observations, it
was realized that such irregular movements of the fluid exhibit distributions
of swirls (eddies, vortices), of various magnitude and vorticities;
it is likely that the large eddies transfer energy to the small eddies,
which dissipate it.\cite{key-2}-\cite{key-4} Statistical concepts
like correlations, homogeneity and isotropy have been introduced in
the theory of turbulence,\cite{key-5,key-6} and dimensional analysis
and similarity arguments allowed the derivation of the energy spectrum
of the turbulent eddies.\cite{key-7}-\cite{key-9}

Meanwhile, the relation of this statistical turbulence with the Navier-Stokes
equation remained unclear.\cite{key-10}-\cite{key-13} Could the
Navier-Stokes equation describe a turbulent motion? To what extent
and in what sense? Has the Navier-Stokes equation smooth and stable
solutions? What is the appropriate representation of a turbulent field
of velocities?\cite{key-14}

The dynamics of the vorticity has enjoyed much interest (see Refs.
\cite{key-15,key-16} and References therein). The results depend
on model assumptions. Dynamical-system concepts and statistical models
have been invoked in studies of turbulence, with chaotic behaviour,
intermittency and coherent structures (see, for instance, Refs. \cite{key-17}-\cite{key-21}).
In particular, by analogy with the quantum turbulence, \textquotedbl Turbulent
flows may be regarded as an intricate collection of mutually-interacting
vortices\textquotedbl , and \textquotedbl Vortex filaments may thus
be seen as the fundamental structure of turbulence,... \textquotedbl .\cite{key-17}

The difficulties exhibited by the Navier-Stokes equation are related
to the viscosity, which governs the vorticity, and the inertial term,
which is quadratic in velocity. We show in this paper that the viscosity
term in the Navier-Stokes equation may produce vorticity, provided
the fluid is incompressible and the flow is isentropic. Although such
an approximate treatment may look reasonable, we can see that, rigorously
speaking, the fluids cannot exhibit vorticity. Moreover, we give arguments
that the vortices are unstable. 

Further, we show in this paper that large variations of the velocity
over short distances lead to highly fluctuating, swirling instabilities,
controlled by viscosity. This is characteristic for the phenomenon
of a fully developed turbulence. In this case, the inertial term acquires
a major role in describing the flow. We represent a fully developed
turbulence as a superposition of fluctuating velocities, associated
to a discrete set of turbulence centres. A mean flow may be included.
In general, the Navier-Stokes equation, averaged over fluctuations,
is not satisfied. On the other hand, a homogeneous and isotropic distribution
of (non-singular) turbulence centres leads to vanishing averages of
velocity and inertial term, such that the Navier-Stokes equation is
satisfied trivially. If the turbulence centres are singular, we are
left with a gas of singularities (or a solution of singularities in
the background fluid), which is in quasi thermal equilibrium. The
corresponding Navier-Stokes equation for the fluid of singularities
is reduced to Newton's equation of motion, with a small friction. 

We illustrate the above descripton with three examples of vorticial
liquids (filamentary liquid, coulombian and dipolar liquid). 

\section{Potential flow. Incompressible fluid}

Let us consider a potential flow of an incompressible fluid. The velocity
$\boldsymbol{v}=grad\Phi$ is given by the gradient of a potential
$\Phi$, which satisfies the Laplace equation 
\begin{equation}
\Delta\Phi=0\label{1}
\end{equation}
 (incompressiblity condition $div\boldsymbol{v}=0$). The viscosity
term $\sim\Delta\boldsymbol{v}$ is zero, such that we are left with
Euler's equation 
\begin{equation}
\frac{\partial\boldsymbol{v}}{\partial t}+(\boldsymbol{v}grad)\boldsymbol{v}=-\frac{1}{\rho}gradp\,\,\,,\label{2}
\end{equation}
 where $\rho$ is the density and $p$ denotes the pressure. By using
the well-known identity 
\begin{equation}
(\boldsymbol{v}grad)\boldsymbol{v}=-\boldsymbol{v}\times curl\boldsymbol{v}+grad(v^{2}/2)\,\,\,,\label{3}
\end{equation}
equation (\ref{2}) becomes 
\begin{equation}
\frac{\partial\boldsymbol{v}}{\partial t}+grad(v^{2}/2+p/\rho)=0\,\,\,,\label{4}
\end{equation}
where $curl\boldsymbol{v}=0$. As it is well known, by using equation
(\ref{1}), this equation leads to 
\begin{equation}
\frac{\partial\Phi}{\partial t}+\frac{1}{2}(grad\Phi)^{2}+\frac{p}{\rho}=0\,\,.\label{5}
\end{equation}
 In this equation $p$ should be viewed as the variation of the pressure
with respect to equilibrium. We assume that $p$ does not depend on
$\Phi$ and the time.

In equations (\ref{1}) and (\ref{5}) the variables may be separated.
Let $g(\boldsymbol{r})$ be a solution of equation (\ref{1}) (satisfying
the boundary conditions); the potential can be written as $\Phi=f(t)g(\boldsymbol{r})$,
where the function $f(t)$ satisfies equation (\ref{5}),
\begin{equation}
\frac{df}{dt}+\frac{1}{2g}(gradg)^{2}+\frac{p}{\rho g}=0\,\,.\label{6}
\end{equation}
 The acceptable solution of this equation (for $f(0)=0$) is
\begin{equation}
f(t)=\frac{\sqrt{2\mid p\mid/\rho}}{\mid gradg\mid}\tanh\frac{\sqrt{\mid p\mid/2\rho}\mid gradg\mid}{g}t\label{7}
\end{equation}
 for $p<0$. It may happen that the boundary conditions for the equation
$\Delta\Phi=0$ depend on time, thus providing the time derivative
$\dot{\Phi}$; in that case equation (\ref{5}) gives the pressure. 

\section{Potential flow. Compressible fluid}

Let us write down the equations of the fluid mechanics
\begin{equation}
\begin{array}{c}
\frac{\partial\rho}{\partial t}+\rho div\boldsymbol{v}+\boldsymbol{v}grad\rho=0\,\,,\\
\\
\rho\frac{\partial\boldsymbol{v}}{\partial t}+\rho(\boldsymbol{v}grad)\boldsymbol{v}=-gradp-\rho grad\varphi+\\
\\
+\eta\Delta\boldsymbol{v}+\left(\frac{1}{3}\eta+\zeta\right)grad\,div\boldsymbol{v}\,\,,\\
\\
\rho T\left(\frac{\partial s}{\partial t}+\boldsymbol{v}grads\right)=\kappa\Delta T+\sigma_{ij}^{'}\partial_{j}v_{i}\,\,\,,
\end{array}\label{8}
\end{equation}
where $p$ is the internal pressure, $\varphi$ is an external potential,
$\eta$ and $\zeta$ are the viscosity coefficients, $T$ is the temperature,
$s$ is the entropy per unit masss, $\kappa$ is the thermoconductivity
and 
\begin{equation}
\sigma_{ij}^{'}=\eta\left(\partial_{i}v_{j}+\partial_{j}v_{i}-\frac{2}{3}\delta_{ij}div\boldsymbol{v}\right)+\zeta\delta_{ij}div\boldsymbol{v}\,\,.\label{9}
\end{equation}
 is the viscosity tensor. In the Navier-Stokes equation (the second
equation (\ref{8})) the forces which determine the velocity are $gradp$
and $\rho grad\varphi$, where $p=p(\rho,T)$ is a function of density
and temperature. For all the usual flows the relative variations $\delta\rho/\rho_{0}$
of the density, $\delta T/T_{0}$ of the temperature, $\delta p/p_{0}$
of the pressure, $\delta s/s_{0}$ of the entropy as well as the variation
$\delta\varphi/\varphi_{0}$ of the external potential are small,
in comparison with their equilibrium values, labelled by the suffix
$0$, when the fluid is at rest. Consequently, we may view the velocity
$\boldsymbol{v}$ as a first-order quantity, and linearize the above
equations as 
\begin{equation}
\begin{array}{c}
\frac{\partial\rho}{\partial t}+\rho_{0}div\boldsymbol{v}=0\,\,,\\
\\
\rho_{0}\frac{\partial\boldsymbol{v}}{\partial t}=-gradp-\rho_{0}grad\varphi+\\
\\
+\eta\Delta\boldsymbol{v}+\left(\frac{1}{3}\eta+\zeta\right)grad\,div\boldsymbol{v}\,\,,\\
\\
\rho_{0}T_{0}\frac{\partial s}{\partial t}=\kappa\Delta T\,\,.
\end{array}\label{10}
\end{equation}
 We note that within this approximation there is no heat source, and
the first-order equation of energy conservation is reduced to an identity. 

The density and entropy variations can be written as 
\begin{equation}
\begin{array}{c}
\delta\rho=\frac{\rho_{0}}{K}\delta p-\beta\rho_{0}\delta T\,\,,\\
\\
\delta s=-\frac{\beta}{\rho_{0}}\delta p+\frac{c_{p}}{T_{0}}\delta T\,\,\,,
\end{array}\label{11}
\end{equation}
where $K$ is the isothermal modulus of compressibility ($1/K=-\frac{1}{V}(\partial V/\partial p)_{T}$,
$V=1/\rho$), $\beta=\frac{1}{V}(\partial V/\partial T)_{p}$ is the
dilatation coefficient and $c_{p}$ is the specific heat per unit
mass at constant pressure, all at equilibrium. In deriving equations
(\ref{11}) the Gibbs free energy $d\Phi=Vdp-sdT$ is used. 

Part of the temperature variation in equation (\ref{11}) is compensated
by pressure variation, as in an adiabatic process; we denote this
contribution by $\delta T_{1}$. The remaining part, denoted by $\delta T$,
corresponds to the conducted heat. Therefore, we write 
\begin{equation}
\begin{array}{c}
\delta\rho=\frac{\rho_{0}}{K}\delta p-\beta\rho_{0}\delta T_{1}-\beta\rho_{0}\delta T\,\,,\\
\\
\delta s=-\frac{\beta}{\rho_{0}}\delta p+\frac{c_{p}}{T_{0}}\delta T_{1}+\frac{c_{p}}{T_{0}}\delta T=\frac{c_{p}}{T_{0}}\delta T\,\,\,,
\end{array}\label{12}
\end{equation}
 whence
\begin{equation}
\frac{\beta}{\rho_{0}}\delta p=\frac{c_{p}}{T_{0}}\delta T_{1}\label{13}
\end{equation}
 and 
\begin{equation}
\delta\rho=\frac{\rho_{0}}{K}\left(1-\frac{\beta^{2}T_{0}K}{\rho_{0}c_{p}}\right)\delta p-\beta\rho_{0}\delta T\,\,.\label{14}
\end{equation}
In this equation we use the thermodynamic relation 
\begin{equation}
\frac{\beta^{2}T_{0}K}{\rho_{0}}=c_{p}-c_{v}\,\,\,,\label{15}
\end{equation}
 where $c_{v}$ is the specific heat per unit mass at constant volume.\cite{key-22}
Therefore, equation (\ref{14}) becomes 
\begin{equation}
\delta\rho=\frac{\rho_{0}c_{v}}{Kc_{p}}\delta p-\beta\rho_{0}\delta T\,\,\,,\label{16}
\end{equation}
 or 
\begin{equation}
\delta p=\frac{Kc_{p}}{\rho_{0}c_{v}}\delta\rho+\frac{\beta Kc_{p}}{c_{v}}\delta T\,\,.\label{17}
\end{equation}
 Now we use two other thermodynamic relations 
\begin{equation}
\frac{c_{p}}{c_{v}}K=K_{ad}\,\,,\,\,\beta K=\alpha\,\,\,,\label{18}
\end{equation}
 where $K_{ad}$ is the adiabatic modulus of compressibility and $\alpha=(\partial p/\partial T)_{V}$
is the thermal pressure coefficient.\cite{key-22} Finally, we get
\begin{equation}
\delta p=\frac{K_{ad}}{\rho_{0}}\delta\rho+\frac{c_{p}}{c_{v}}\alpha\delta T\,\,\,,\label{19}
\end{equation}
 which is used in the Navier-Stokes equation. Equations (\ref{10})
become 
\begin{equation}
\begin{array}{c}
\frac{\partial\rho}{\partial t}+\rho_{0}div\boldsymbol{v}=0\,\,,\\
\\
\rho_{0}\frac{\partial\boldsymbol{v}}{\partial t}=-\frac{K_{ad}}{\rho_{0}}grad\rho-\rho_{0}grad\varphi-\frac{c_{p}}{c_{v}}\alpha gradT+\\
\\
+\eta\Delta\boldsymbol{v}+\left(\frac{1}{3}\eta+\zeta\right)grad\,div\boldsymbol{v}\,\,,\\
\\
\rho_{0}c_{p}\frac{\partial T}{\partial t}=\kappa\Delta T\,\,\,,
\end{array}\label{20}
\end{equation}
where the second equation (\ref{12}) is used. We can see that the
temperature equation (\ref{20}) is independent; it describes the
transport of an external temperature, which may provide a source for
the velocity in the Navier-Stokes equation. We may leave aside this
external temperature. 

Let us seek a potential-flow solution of the above equations, where
the velocity is derived from a potential $\Phi$, by 
\begin{equation}
\boldsymbol{v}=grad\Phi\,\,.\label{21}
\end{equation}
 We notice that $curl\boldsymbol{v}=0$ and $curl\,curl\boldsymbol{v}=0$,
\emph{i.e.} $\Delta\boldsymbol{v}=grad\,div\boldsymbol{v}$. Therefore,
the Navier-Stokes equation can be written as 
\begin{equation}
\begin{array}{c}
\rho_{0}\frac{\partial\boldsymbol{v}}{\partial t}=-\frac{K_{ad}}{\rho_{0}}grad\rho-\rho_{0}grad\varphi+\\
\\
+\left(\frac{4}{3}\eta+\zeta\right)grad\,div\boldsymbol{v}\,\,.
\end{array}\label{22}
\end{equation}
 By using equation (\ref{21}), we obtain 
\begin{equation}
\begin{array}{c}
\frac{\partial\rho}{\partial t}+\rho_{0}\Delta\Phi=0\,\,,\\
\\
\frac{\partial\Phi}{\partial t}+\frac{K_{ad}}{\rho_{0}^{2}}\rho+\varphi-\frac{1}{\rho_{0}}\left(\frac{4}{3}\eta+\zeta\right)\Delta\Phi=0\,\,\,,
\end{array}\label{23}
\end{equation}
up to a function of time, where $\rho$ and $\varphi$ should be viewed
as their corresponding variations. By an additional time differentiation
we obtain
\begin{equation}
\frac{\partial^{2}\Phi}{\partial t^{2}}-\frac{K_{ad}}{\rho_{0}}\Delta\Phi+\dot{\varphi}-\frac{1}{\rho_{0}}\left(\frac{4}{3}\eta+\zeta\right)\Delta\dot{\Phi}=0\,\,.\label{24}
\end{equation}
 This equation provides the potential $\Phi$, therefore the velocity
$\boldsymbol{v}$ through equation (\ref{21}) and the density $\rho$
through the first equation (\ref{23}).

Equation (\ref{24}) is the wave equation with friction (the term
$\sim\Delta\dot{\Phi}$) and sources ($-\dot{\varphi}$). The ratio
$K_{ad}/\rho_{0}$ is the square of the sound velocity $c=\sqrt{K_{ad}/\rho_{0}}$.
The elementary solutions $e^{-i\omega t}e^{i\boldsymbol{kr}}$ of
this (homogeneous) equation are damped plane waves
\begin{equation}
e^{\mp ickt}e^{i\boldsymbol{kr}}e^{-\frac{\sigma k^{2}}{c}t}\,\,\,,\label{25}
\end{equation}
 for $\sigma k\ll c$, where $\sigma=(4\eta/3+\zeta)/2\rho_{0}$.
The relaxation time is much longer than the wave period. A wave propagating
along the $x$-direction is proportional to $\sim e^{-\gamma x}$,
with the attenuation coefficient $\gamma=\sigma k^{2}/c=\sigma\omega^{2}/c^{3}$.
This is the well-known absorption coefficient for sound (without the
$\kappa$-contribution). 

\section{Vorticity}

Euler's equation for an ideal fluid can be written as 
\begin{equation}
\frac{d\boldsymbol{v}}{dt}=-gradw\,\,\,,\label{26}
\end{equation}
 where $w$ is the enthalpy ($dw=\frac{1}{\rho}dp$); the pressure
$p$ is a function of density $\rho$. By taking the $curl$, we get
\begin{equation}
curl\frac{d\boldsymbol{v}}{dt}=0\,\,.\label{27}
\end{equation}
 On the other hand, 
\begin{equation}
\frac{d\boldsymbol{v}}{dt}=\frac{\partial\boldsymbol{v}}{\partial t}+(\boldsymbol{v}grad)\boldsymbol{v}=\frac{\partial\boldsymbol{v}}{\partial t}+\left(\frac{\partial\boldsymbol{v}}{\partial t}\right)_{f}\,\,\,,\label{28}
\end{equation}
 where the suffix $f$ indicates that the derivative is taken along
the flow. Equation (\ref{27}) becomes
\begin{equation}
\frac{\partial}{\partial t}curl\boldsymbol{v}+\left(\frac{\partial}{\partial t}curl\boldsymbol{v}\right)_{f}=0\,\,;\label{29}
\end{equation}
since the two variations of the $curl\boldsymbol{v}$ are independent,
we get 
\begin{equation}
curl\boldsymbol{v}=0\,\,\,,\label{30}
\end{equation}
\emph{ i.e.} the vorticity $curl\boldsymbol{v}$ is conserved along
the flow. Therefore, we cannot create, or destroy, vorticity $curl\boldsymbol{v}$
in the flow of an ideal fluid. Equation (\ref{30}) is valid in the
absence of special external force, which do not derive from a gradient.
This is Helmholtz's circulation law. As it is well known, an ideal
fluid supports only an irrotational (potential) flow, where the velocity
is derived from a scalar potential ($\boldsymbol{v}=grad\Phi$). By
knowing the equation of state of the fluid, the Euler equation and
the continuity equation are fully determined. 

For a real, viscid, fluid the Navier-Stokes equation is 
\begin{equation}
\rho\frac{d\boldsymbol{v}}{dt}=-gradp+\eta\Delta\boldsymbol{v}+\left(\frac{1}{3}\eta+\zeta\right)grad\,div\boldsymbol{v}\,\,\,,\label{31}
\end{equation}
 where $\eta$, $\zeta$ are the viscosity coefficients. By taking
the $curl$, we get 
\begin{equation}
\begin{array}{c}
curl\left(\rho\frac{d\boldsymbol{v}}{dt}\right)=grad\rho\times\frac{d\boldsymbol{v}}{dt}+\rho curl\frac{d\boldsymbol{v}}{dt}=\\
\\
=\eta\Delta curl\boldsymbol{v}\,\,;
\end{array}\label{32}
\end{equation}
we can see that the viscosity $\eta$ can generate vorticity ($curl\boldsymbol{v}\neq0$).
In general, the velocity $\boldsymbol{v}$ rotates about the vorticity
$curl\boldsymbol{v}$. This is a vortex. 

Equation (\ref{32}) is in conflict with the continuity equation 
\begin{equation}
\frac{d\rho}{dt}+\rho div\boldsymbol{v}=0\,\,.\label{33}
\end{equation}
 Indeed, if $curl\boldsymbol{v}\neq0$, the velocity should be derived
from a $curl$ (not from a $grad$!), \emph{i.e.} we should have $\boldsymbol{v}=curl\boldsymbol{A}$,
where $\boldsymbol{A}$ is a vector potential. Consequently, the fluid
should be incompressible ($div\boldsymbol{v}=div\,curl\boldsymbol{A}=0$).
The density should be constant, both in time and space (along the
flow). This indicates that in a compressible fluid we cannot have
vortices. Usually, the variations of the density are small, such that
they may be neglected for the present purpose. 

Therefore, we may limit to an incompressible fluid ($div\boldsymbol{v}=0$),
for which equation (\ref{32}) becomes 
\begin{equation}
curl\frac{d\boldsymbol{v}}{dt}=\nu\Delta curl\boldsymbol{v}\,\,\,,\label{34}
\end{equation}
 or 
\begin{equation}
\frac{\partial}{\partial t}curl\boldsymbol{v}-curl\left(\boldsymbol{v}\times curl\boldsymbol{v}\right)=\nu\Delta curl\boldsymbol{v}\,\,\,,\label{35}
\end{equation}
where $\nu=\eta/\rho$ is the kinematical viscosity and we have used
the identity $(\boldsymbol{v}grad)\boldsymbol{v}=-\boldsymbol{v}\times curl\boldsymbol{v}+grad(v^{2}/2)$.
This is the equation of vorticity; it can also be written as 

\begin{equation}
\frac{\partial}{\partial t}curl\boldsymbol{v}+curl\left[(\boldsymbol{v}grad)\boldsymbol{v}\right]=\nu\Delta curl\boldsymbol{v}\,\,.\label{36}
\end{equation}

This equation gives the velocity. The pressure is obtained from the
Navier-Stokes equation. If we write the Navier-Stokes equation as
\begin{equation}
\frac{\partial}{\partial t}curl\boldsymbol{A}+(\boldsymbol{v}grad)\boldsymbol{v}=-\frac{1}{\rho}gradp+\nu\Delta curl\boldsymbol{A}\,\,\,,\label{37}
\end{equation}
we get 
\begin{equation}
\begin{array}{c}
div\left[(\boldsymbol{v}grad)\boldsymbol{v}+\frac{1}{\rho}gradp\right]=0\end{array}\label{38}
\end{equation}
 and 
\begin{equation}
div\left(\frac{\partial\boldsymbol{v}}{\partial t}-\nu\Delta\boldsymbol{v}\right)=0\,\,\,,\label{39}
\end{equation}
 which is an identity ($div\boldsymbol{v}=0$, $\boldsymbol{v}=curl\boldsymbol{A}$).
In some cases a particular solution of these equations is provided
by 
\begin{equation}
\begin{array}{c}
(\boldsymbol{v}grad)\boldsymbol{v}=-grad(p/\rho)\,\,,\\
\\
\frac{\partial\boldsymbol{v}}{\partial t}=\nu\Delta\boldsymbol{v}\,\,.
\end{array}\label{40}
\end{equation}
We can see that the Navier-Stokes equation is split into (the derivatives
of) a diffusion (heat) equation and an equilibrium equation; the first
equation (\ref{40}) indicates an equilibrium between the pressure
force $-grad(p/\rho)$ and Euler's force $(\boldsymbol{v}grad)\boldsymbol{v}$.
The diffusion equation (\ref{40}) holds also for the vorticity, because
the above equations are valid for a non-vanishing vorticity. It is
easy to see that these equations generalize the equations for the
Couette flow. 

However, we have also the heat-transfer equation. For an incompressible
fluid it reads
\begin{equation}
\rho c_{p}\frac{dT}{dt}=\kappa\Delta T+\frac{1}{2}\eta\left(\partial_{i}v_{j}+\partial_{j}v_{i}\right)^{2}\,\,\,,\label{41}
\end{equation}
or 
\begin{equation}
\frac{dT}{dt}=\chi\Delta T+\frac{1}{2}\frac{\nu}{c_{p}}\left(\partial_{i}v_{j}+\partial_{j}v_{i}\right)^{2}\,\,\,,\label{42}
\end{equation}
 where $c_{p}$ is the specific heat per unit mass at constant pressure,
$\kappa$ is the thermoconductivity, and $\chi=\kappa/\rho c_{p}$
is the thermometric conductivity. For an incompressible fluid this
equation can be transformed into an equation for the derivatives of
the pressure, 
\begin{equation}
\frac{dp}{dt}=\chi\Delta p+\frac{1}{2}\frac{\alpha\nu}{c_{p}}\left(\partial_{i}v_{j}+\partial_{j}v_{i}\right)^{2}\,\,\,,\label{43}
\end{equation}
 where $\alpha=(\partial p/\partial T)_{v}$ is the thermal pressure
coefficient. In general, equation (\ref{43}) is not compatible with
the Navier-Stokes equation. Therefore, rigorously speaking, we cannot
have vorticity in an incompressible fluid either. Usually, the coefficient
$\nu/c_{p}$ is very small (of the order $10^{-24}-10^{-25}g\cdot cm^{2}/s$),
such that, for small gradients of velocity, we have a low rate of
entropy production (though, a factor of the order $10^{23}K/erg$
should be taken into account). Under these conditions, we may assume
that the flow is isentropic and the heat-transfer equation may be
neglected. 

In order to get an idea of how large the variations of the density
and the temperature can be, we may estimate a change $\delta p$ in
pressure from $\delta p\simeq\rho v^{2}$. A velocity $v=100km/h$,
which is fairly large, produces a change $\delta p\simeq10^{4}dyn/cm^{2}$
in air ($\rho=10^{-3}g/cm^{3}$), whose normal pressure is $10^{6}dyn/cm^{2}$;
therefore, $\delta p/p\simeq10^{-2}$. Such a velocity ($\simeq3\times10^{3}cm/s$)
is close to the mean thermal velocity $\simeq10^{4}cm/s$ (for normal
air), and close to the sound velocity in normal air $c\simeq3.5\times10^{4}cm/s$.
For this velocity we still expect local thermal equilibrium. The change
in density is given by $\delta p=K(\delta\rho/\rho)$, where $K=-V(\partial p/\partial V)$
is the (say, isothermal) modulus of compression. For air $K\simeq10^{6}dyn/cm^{2}$,
for water $K\simeq10^{10}dyn/cm^{2}$, such that we get $\delta\rho/\rho\simeq10^{-2},\,10^{-6}$.
The change in temperature is obtained from $\delta p=\alpha T(\delta T/T)$,
where $\alpha=(\partial p/\partial T)_{V}$ is the thermal pressure
coefficient (at constant volume $V$). For water $\alpha\simeq10^{22}/cm^{3}$,
for gases it is much higher; for normal temperature $T=300K$ we get
$\delta T/T\simeq10^{-4}$, or much lower. Consequently, we may expect
an almost ideal, incompressible flow. We note that, although we neglect
the viscosity in the heat-transfer equation, we keep it in the Navier-Stokes
equation. 

Therefore, within these approximations (incompressibility and constant
entropy), we are left with equation (\ref{36}) and the Navier-Stokes
equation for a vorticial flow. In general, an external pressure which
satisfies the Navier-Stokes equation (or equation (\ref{38})) is
very special, such that, if they exist, the vortices might be, in
fact, unstable. They develop an Euler's force, which is difficult
to be compensated by an external force. 

We note that, under the conditions stated above, the viscosity may
generate (unstable) vortices. In the next section we show that the
viscosity may generate another type of instabilities. 

For small variations of the velocity we may neglect the inertial term
in the vorticity equation (\ref{36}), which becomes
\begin{equation}
\frac{\partial}{\partial t}curl\boldsymbol{v}=\nu\Delta curl\boldsymbol{v}\,\,.\label{44}
\end{equation}
 By making use of $\boldsymbol{v}=curl\boldsymbol{A}$, we get 
\begin{equation}
\left(\Delta-grad\,div\right)\left(\frac{\partial\boldsymbol{A}}{\partial t}-\nu\Delta\boldsymbol{A}\right)=0\,\,.\label{45}
\end{equation}
A solution of this equation is provided by 
\begin{equation}
\frac{\partial\boldsymbol{A}}{\partial t}-\nu\Delta\boldsymbol{A}=0\,\,\,,\label{46}
\end{equation}
 which leads to 
\begin{equation}
\boldsymbol{A}=\boldsymbol{A}_{0}e^{-\lambda\nu t}e^{\pm i\sqrt{\lambda}r}/r\,\,\,,\label{47}
\end{equation}
 where $\boldsymbol{A_{0}}$ and $\lambda$ are two constants. The
velocity acquires the form 
\begin{equation}
\boldsymbol{v}=-\boldsymbol{A}_{0}\times grad\left(e^{-\lambda\nu t}e^{\pm i\sqrt{\lambda}r}/r\right)\,\,\,,\label{48}
\end{equation}
 and the pressure is uniform within this approximation. We note that,
although the spatial dependence of the solution does not depend on
viscosity, it is generated by the viscosity term $\nu\Delta\boldsymbol{v}$.

\section{Instabilities}

The equation of energy conservation for an incompressible fluid is
\begin{equation}
\begin{array}{c}
\frac{\partial}{\partial t}\left(\frac{1}{2}\rho v^{2}\right)+div\left[\boldsymbol{v}\left(\frac{1}{2}\rho v^{2}+p\right)-\frac{1}{2}\eta grad(v^{2})\right]+\\
\\
+\eta\left(\partial_{j}v_{i}\right)^{2}=0\,\,;
\end{array}\label{49}
\end{equation}
it is obtained by multiplying by $\boldsymbol{v}$ the Navier Stokes
equation (\ref{31}) for an incompressible fluid ($div\boldsymbol{v}=0$).
The $div$-term represents a transport of energy and mechanical work
of the pressure, and an energy flux associated with collisions (viscosity);
the term $\eta\left(\partial_{j}v_{i}\right)^{2}$ represents the
heat produced by viscosity. We integrate this equation over a volume
$V$ enclosed by a surface $S$, 
\begin{equation}
\begin{array}{c}
\frac{\partial}{\partial t}\int dV\left(\frac{1}{2}\rho v^{2}\right)+\oint dS\left[v_{n}\left(\frac{1}{2}\rho v^{2}+p\right)-\frac{1}{2}\eta\partial_{n}(v^{2})\right]+\\
\\
+\eta\int dV\left(\partial_{j}v_{i}\right)^{2}=0\,\,\,,
\end{array}\label{50}
\end{equation}
 where $v_{n}$ is the velocity component normal to the surface and
$\partial_{n}$ is the derivative along the normal to the surface. 

We compare the orders of magnitude of the surface terms and the $\eta$-volume
term, and get ratios of the form $\frac{Sl}{V}R$, $\frac{Sl}{V}(p/\rho v^{2})R$,
$\frac{Sl}{V}$, where $l$ is the distance over which the velocity
varies and $R=vl/\nu$ is the Reynolds number. For moderate Reynolds
numbers and $Sl/V\ll1$ we can neglect the surface contributions in
comparison with the heat term. By writing 
\begin{equation}
\boldsymbol{v}=f(t)\boldsymbol{u}(\boldsymbol{r})\,\,\,,\label{51}
\end{equation}
the above equation becomes 
\begin{equation}
\frac{\partial}{\partial t}f^{2}\cdot\int dV\left(\frac{1}{2}\rho u^{2}\right)+\eta f^{2}\int dV\left(\partial_{j}u_{i}\right)^{2}=0\,\,.\label{52}
\end{equation}
 We can see that the time dependence of the velocity is a damped exponential.
The flow is stable, as a consequence of the dissipated heat. The $\eta$-term
in equation (\ref{52}) gives, in fact, the increase of entropy. 

Let us assume that the integration domains are sufficiently small,
such that $Sl/V$ is of the order of unity; the velocity varies over
a distance $l$ inside the domains, but we assume that it suffers
a large discontinuity across the surface, over a small distance $\delta\ll l$.
Then, it is easy to see that the dominant term in equation (\ref{50})
is the collision term, such that equation (\ref{50}) becomes 
\begin{equation}
\frac{\partial}{\partial t}f^{2}\cdot\int dV\left(\frac{1}{2}\rho u^{2}\right)-\frac{1}{2}\eta f^{2}\oint dS\partial_{n}(u^{2})=0\,\,.\label{53}
\end{equation}
 We can see that for a positive normal derivative the flow is unstable.
The viscosity is insufficient to disipate the energy as heat, and
the energy is transferred by molecular collisions (viscosity) through
surfaces of discontinuities. The process occurs in small domains,
with large discontinuities of velocity across their surface, and the
instabilities imply returning, swirling and fluctuating, velocities.
This is the turbulence phenomenon. We note that the instabilities
are governed by viscosity, which gives also vorticity (when the entropy
production is neglected). Moreover, we note that the inertial term
does not appear in instabilities, though it plays an important role
in turbulence. 

The above arguments can be extended to compressible fluids, including
the variations of the temperature. Indeed, the energy conservation
in this case reads
\begin{equation}
\frac{\partial}{\partial t}\left(\frac{1}{2}\rho v^{2}+\rho\varepsilon\right)=-\partial_{j}\left[\rho v_{j}\left(\frac{1}{2}v^{2}+w\right)-v_{i}\sigma_{ij}^{'}-\kappa\partial_{j}T\right]\,\,\,,\label{54}
\end{equation}
 where $\varepsilon$ is the internal energy per unit mass, $w$ is
the enthalpy per unit mass and 
\begin{equation}
\sigma_{ij}^{'}=\eta\left(\partial_{i}v_{j}+\partial_{j}v_{i}-\frac{2}{3}\delta_{ij}div\boldsymbol{v}\right)+\zeta\delta_{ij}div\boldsymbol{v}\label{55}
\end{equation}
 is the viscosity tensor. For a smooth flow and a sufficiently large
volume the surface term in equation (\ref{54}) can be neglected,
and the energy is conserved, as it is well known. However, in the
surface integral we have terms of the form 
\begin{equation}
\oint dS\left[\eta v_{i}\left(\partial_{i}v_{n}+\partial_{n}v_{i}\right)-\left(\frac{2}{3}\eta-\zeta\right)v_{n}div\boldsymbol{v}+\kappa\partial_{n}T\right]\,\,\,,\label{56}
\end{equation}
 which imply normal derivatives to the surface, both of velocity and
temperature. By collecting these contributions, we get 
\begin{equation}
\oint dS\left[\eta\partial_{n}(v^{2}/2)+\left(\frac{1}{3}\eta+\zeta\right)\partial_{n}(v_{n}^{2}/2)+\kappa\partial_{n}T\right]\,\,.\label{57}
\end{equation}
 We can see that for large normal derivatives across the surface,
both for velocity and temperature, these terms may lead to instabilities.

\section{Turbulence}

As it is well known, for a moderate turbulence, \emph{i.e.} for slowly
varying fluctuations, we may decompose the velocity field into a mean
velocity and a fluctuating part, and limit ourselves to the time averaged
Navier-Stokes equation. This way we get the Reynolds equations, for
which the mean energy is coupled to the fluctuating energy, via model
assumptions. A fully developed turbulence exhibits highly-varying
fluctuations, such that we need to consider the time-dependent Navier-Stokes
equation. 

The turbulent instabilities ocurring in a fully developed turbulence
exhibit large variations of the velocity over small distances. In
this case we may assume that the velocity is split into a mean-flow
velocity $\boldsymbol{v}_{0}$ and a fluctuating part $\boldsymbol{v}$,
where the mean-flow velocity $\boldsymbol{v}_{0}$ may have a slight
time variation, while the fluctuating velocity $\boldsymbol{v}$ is
a rapidly varying velocity. By using this decomposition for an incompressible
fluid, the Navier-Stokes equation reads 
\begin{equation}
\begin{array}{c}
\frac{\partial\boldsymbol{v}_{0}}{\partial t}+\frac{\partial\boldsymbol{v}}{\partial t}+\left(\boldsymbol{v}_{0}grad\right)\boldsymbol{v}_{0}+\left(\boldsymbol{v}_{0}grad\right)\boldsymbol{v}+\left(\boldsymbol{v}grad\right)\boldsymbol{v}_{0}+\\
\\
+\left(\boldsymbol{v}grad\right)\boldsymbol{v}=-\frac{1}{\rho}gradp_{0}-\frac{1}{\rho}gradp+\nu\Delta\boldsymbol{v}_{0}+\nu\Delta\boldsymbol{v}\,\,\,,
\end{array}\label{58}
\end{equation}
 where $p_{0}$ is the pressure corresponding to the main flow and
$p$ is the fluctuating part of the pressure. In this equation we
have three distinct types of time variations, such that it should
be viewed as three equations
\begin{equation}
\begin{array}{c}
\frac{\partial\boldsymbol{v}_{0}}{\partial t}+\left(\boldsymbol{v}_{0}grad\right)\boldsymbol{v}_{0}=-\frac{1}{\rho}gradp_{0}+\nu\Delta\boldsymbol{v}_{0}\,\,,\\
\\
\frac{\partial\boldsymbol{v}}{\partial t}+\left(\boldsymbol{v}_{0}grad\right)\boldsymbol{v}+\left(\boldsymbol{v}grad\right)\boldsymbol{v}_{0}=-\frac{1}{\rho}gradp+\nu\Delta\boldsymbol{v}\,\,,\\
\\
\left(\boldsymbol{v}grad\right)\boldsymbol{v}=0\,\,;
\end{array}\label{59}
\end{equation}
similarly, the continuity equation should be split into 
\begin{equation}
div\boldsymbol{v}_{0}=0\,\,,\,\,div\boldsymbol{v}=0\,\,.\label{60}
\end{equation}

The first equation (\ref{59}) is an independent equation, which gives
the main flow velocity $\boldsymbol{v}_{0}$. Since the main part
of the velocity is taken by the fluctuating velocity, we may neglect
the quadratic term in this equation. The energy conservation and the
heat transfer for this equation are given by 
\begin{equation}
\begin{array}{c}
\frac{\partial}{\partial t}(v_{0}^{2}/2)=-\partial_{i}\left[v_{0i}\left(p_{0}/\rho+v_{0}^{2}/2\right)-\nu\partial_{i}(v_{0}^{2}/2)\right]-\\
\\
-\nu(\partial_{i}v_{0j})^{2}\,\,,\\
\\
T_{0}\frac{ds_{0}}{dt}=\chi\Delta T_{0}+\nu(\partial_{i}v_{0j})^{2}\,\,\,,
\end{array}\label{61}
\end{equation}
 where $T_{0}$ is the temperature of the main flow, $s_{0}$ is the
entropy per unit mass of the main flow and $\chi$ is the thermometric
conductivity.

Having solved the mean-flow equation we can pass to solve the second
equation (\ref{59}) for the fluctuating velocity $\boldsymbol{v}$,
with $\boldsymbol{v}_{0}$ as a parameter. This equation has its own
energy-conservation and heat-transfer equations. We note that the
temperature and the entropy of the mean flow are different from the
temperature and the entropy of the fluctuating part of the flow, which
means that the two components of the flow (the mean flow and the fluctuating
flow) are not in thermal equilbrium. Indeed, if we multiply the first
equation (\ref{59}) by $\boldsymbol{v}$ and the second equation
(\ref{59}) by $\boldsymbol{v}_{0}$, we get cross-terms of the form
$T_{0}\frac{ds}{dt}+T\frac{ds_{0}}{dt}$ in the heat-transfer equation,
where $T$ and $s$ are the temperature and the entropy of the fluctuating
flow. This indicates a heat exchange between the two components of
the flow.

We are left with the third equation (\ref{59}), which, in general,
is not satisfied. We conclude that the fully developed turbulence
does not satisfy the Navier-Stokes equation. The quadratic term of
the third equation (\ref{59}) is equivalent to a rapidly varying
internal force (Euler's force), which cannot be compensated by any
physical external force. The fully developed turbulence is unstable.

Under these conditions it is reasonable to be interested in time averaged
quantities. Then, the fluctuating part of the flow is reduced to 
\begin{equation}
\overline{\left(\boldsymbol{v}grad\right)\boldsymbol{v}}=0\,\,.\label{62}
\end{equation}
Since $div\boldsymbol{v}=0$, the components of the velocity $\boldsymbol{v}$
are not independent. In general, equation (\ref{62}) is not satisfied,
which means that the turbulent motion is unstable even on average.
We note that a similar decomposition is valid for a compressible fluid,
as long as the velocity, density and entropy fluctuations are independent
of one another. 

Since a fully developed turbulence originates in large variations
of the velocity across small distances, it is reasonable to associate
these variations with a discrete distribution of positions $\boldsymbol{r}_{i},$which
we call centres of turbulence. Further on, we assume that this is
a homogeneous and isotropic distribution, such that we may write the
velocity field as 
\begin{equation}
\boldsymbol{v}=\sum_{i}\boldsymbol{v}_{i}(t,R_{i})\,\,\,,\label{63}
\end{equation}
 where $\boldsymbol{R}_{i}=\boldsymbol{r}-\boldsymbol{r}_{i}$. If
$curl\boldsymbol{v}_{i}\neq0$, this velocity field represents a vorticial
liquid, which is unstable. We assume that the fluctuating velocities
are independent at distinct positions, \emph{i.e.} $\overline{\boldsymbol{v}_{i}}=0$
and $\overline{\boldsymbol{v}_{i}\boldsymbol{v}_{j}}\sim\delta_{ij}$.
Equation (\ref{62}) becomes 
\begin{equation}
\overline{\left(\boldsymbol{v}grad\right)\boldsymbol{v}}=\sum_{i}\overline{\left(\boldsymbol{v}_{i}\boldsymbol{R}_{i}/R_{i}\right)\left(d\boldsymbol{v}_{i}/dR_{i}\right)}\,\,.\label{64}
\end{equation}
The conditions of homogeneity and isotropy imply that $\boldsymbol{v}_{i}$
in the above equation may be replaced by the same velociy $\boldsymbol{u}$.
For a sufficiently dense set of positions $\boldsymbol{r}_{i}$ we
can define a density $\rho_{v}$ of such points, which is a constant.
Then, equation (\ref{64}) can be transformed into the integral 
\begin{equation}
\overline{\left(\boldsymbol{v}grad\right)\boldsymbol{v}}=\rho_{v}\int dR\cdot R^{2}\int do\overline{\left(\boldsymbol{u}\boldsymbol{R}/R\right)\left(d\boldsymbol{u}/dR\right)}\,\,.\label{65}
\end{equation}
If the radial integral is finite, the result of integration in the
above equation is zero, due to the integration over the solid angle
$o$, such that the term $\overline{\left(\boldsymbol{v}grad\right)\boldsymbol{v}}$
is zero. In this case we can say that the Navier-Stokes equation is
trivially satisfied on average, being reduced to the mean-flow equation
(first equation (\ref{59})). 

If the radial integral in equation (\ref{65}) is singular for $\boldsymbol{R}=0$,
as it may often happen for vortices, we are left with a discrete set
of singularities, extending over a small characteristic distance $a$,
where the singularity is 
\begin{equation}
\boldsymbol{u}\sim(a/R)^{n}\,\,,\,\,n>1\,\,.\label{66}
\end{equation}
In each of these regions there exists a mass $M$ of fluid, which
can be carried by the background fluid and, at the same time, they
may have their own motion.

The averaged energy-conservation equation derived from the second
equation (\ref{59}), 
\begin{equation}
\frac{1}{2}\boldsymbol{v}_{0}grad\overline{v^{2}}+\partial_{j}\left(\overline{v_{i}v_{j}}v_{0i}\right)-\frac{1}{2}\nu\Delta\overline{v^{2}}=-\nu\overline{\left(\partial_{j}v_{i}\right)^{2}}\,\,\,,\label{67}
\end{equation}
 shows that the fluctuating motion produces heat which is partly transported
by the mean flow (the $div$-terms integrated over a volume are irrelevant).
This dissipated heat should be compensated from the outside. Therefore,
we are left with a quasi ideal classical gas of singular vortices,
or a solution of vortices in the background fluid, in thermal quasi
equilibrium. This is an example of emergent dynamics.\cite{key-23}

\section{Gas of singularities}

Let us assume a homogeneous, isotropic, fluctuating distribution of
singular centres of turbulence localized at $\boldsymbol{r}_{i}$
with mass $M$, as described above. Their density is 
\begin{equation}
\rho=M\sum_{i}\delta(\boldsymbol{r}-\boldsymbol{r}_{i})\label{68}
\end{equation}
 and 
\begin{equation}
\frac{\partial\rho}{\partial t}=-M\sum_{i}\boldsymbol{u}_{i}grad\delta(\boldsymbol{r}-\boldsymbol{r}_{i})\,\,\,,\label{69}
\end{equation}
where $\boldsymbol{u}_{i}=d\boldsymbol{r}_{i}/dt$ is their velocity.
The velocity field of the singularities is 
\begin{equation}
\boldsymbol{u}=v\sum_{i}\boldsymbol{u}_{i}\delta(\boldsymbol{r}-\boldsymbol{r}_{i})\,\,\,,\label{70}
\end{equation}
 where $v$ is the small volume over which the $\delta$-function
is localized, such that $v=a^{3}$ and $M=\rho v$. Let us compute
\begin{equation}
\begin{array}{c}
div(\rho\boldsymbol{u})=vMdiv\sum_{ij}\delta(\boldsymbol{r}-\boldsymbol{r}_{i})\boldsymbol{u}_{j}\delta(\boldsymbol{r}-\boldsymbol{r}_{j})=\\
\\
=Mdiv\sum_{i}\boldsymbol{u}_{i}\delta(\boldsymbol{r}-\boldsymbol{r}_{i})=\\
\\
=M\sum_{i}\boldsymbol{u}_{i}grad\delta(\boldsymbol{r}-\boldsymbol{r}_{i})=-\frac{\partial\rho}{\partial t}\,\,;
\end{array}\label{71}
\end{equation}
we can see that the continuity equation is satisfied. 

Now, let us focus on the Navier-Stokes equation 
\begin{equation}
\rho\frac{\partial\boldsymbol{u}}{\partial t}+\rho(\boldsymbol{u}grad)\boldsymbol{u}=-gradp+\eta\Delta\boldsymbol{u}\,\,,\label{72}
\end{equation}
and let us compute each term in this equation for our fluid of singularities.
We have 
\begin{equation}
\begin{array}{c}
\frac{\partial\boldsymbol{u}}{\partial t}=v\sum_{i}\dot{\boldsymbol{u}}_{i}\delta(\boldsymbol{r}-\boldsymbol{r}_{i})-v\sum_{i}\boldsymbol{u}_{i}(\boldsymbol{u}_{i}grad)\delta(\boldsymbol{r}-\boldsymbol{r}_{i})\,\,.\end{array}\label{73}
\end{equation}
 The inertial term is 
\begin{equation}
\begin{array}{c}
(\boldsymbol{u}grad)\boldsymbol{u}=v^{2}\sum_{ij}\boldsymbol{u}_{j}\delta(\boldsymbol{r}-\boldsymbol{r}_{i})(\boldsymbol{u}_{i}grad)\delta(\boldsymbol{r}-\boldsymbol{r}_{j})=\\
\\
=v\sum_{i}\boldsymbol{u}_{i}(\boldsymbol{u}_{i}grad)\delta(\boldsymbol{r}-\boldsymbol{r}_{i})\,\,.
\end{array}\label{74}
\end{equation}
 On comparing equations (\ref{73}) and (\ref{74}), we can see that
the inertial (Euler's) term disappears from equation. This is expected,
since the $\delta$-function equals the variable $\boldsymbol{r}$
to the function $\boldsymbol{r}_{i}(t)$, which amounts to Lagrange's
approach. The term on the left in equation (\ref{72}) becomes
\begin{equation}
\begin{array}{c}
\rho\frac{\partial\boldsymbol{u}}{\partial t}+\rho(\boldsymbol{u}grad)\boldsymbol{u}=M\sum_{i}\dot{\boldsymbol{u}}_{i}\delta(\boldsymbol{r}-\boldsymbol{r}_{i})\,\,\,,\end{array}\label{75}
\end{equation}
and equation (\ref{72}) reads now 
\begin{equation}
\begin{array}{c}
M\sum_{i}\dot{\boldsymbol{u}}_{i}\delta(\boldsymbol{r}-\boldsymbol{r}_{i})=-v\sum_{i}p_{i}grad\delta(\boldsymbol{r}-\boldsymbol{r}_{i})+\\
\\
+\eta v\sum_{i}\boldsymbol{u}_{i}\Delta\delta(\boldsymbol{r}-\boldsymbol{r}_{i})\,\,\,,
\end{array}\label{76}
\end{equation}
 where $p_{i}$ is the pressure at the position $\boldsymbol{r}_{i}$.
The pressure term in equation (\ref{76}) is a force per unit volume
acting upon the vortex placed at $\boldsymbol{r}_{i}$; it may arise
from the pressure exerted by the background fluid particles. This
term may be written as 
\begin{equation}
\sum_{i}\boldsymbol{f}_{i}\delta(\boldsymbol{r}-\boldsymbol{r}_{i})\,\,\,,\label{77}
\end{equation}
where $\boldsymbol{f}_{i}$ is the force acting at $\boldsymbol{r}_{i}$.
The factor $\Delta\delta(\boldsymbol{r}-\boldsymbol{r}_{i})$ is of
the order $-\frac{1}{a^{2}}\delta(\boldsymbol{r}-\boldsymbol{r}_{i})$,
where $a$ is of the order of the dimension of the vortex ($v=a^{3})$;
consequently, we may replace the viscosity terms in equation (\ref{76})
by 
\begin{equation}
-\frac{\eta v}{a^{2}}\sum_{i}\boldsymbol{u}_{i}\delta(\boldsymbol{r}-\boldsymbol{r}_{i})\,\,.\label{78}
\end{equation}
A similar contribution brings the $\zeta$-term. The equation of motion
(\ref{76}) describes a set of independent particles with mass $M$,
subjected to an external force $\boldsymbol{f}_{i}$ and a friction
force; the equation of motion of each such particle can be written
as 
\begin{equation}
M\dot{\boldsymbol{u}}_{i}=\boldsymbol{f}_{i}-\eta a\boldsymbol{u}_{i}\,\,\,,\label{79}
\end{equation}
which is Newton's law of motion. The damping coefficient caused by
the friction force is very small, such that we can consider the ensemble
of singularities as a (quasi) ideal classical gas of independent,
identical, pointlike particles. Therefore, a (singular) fully developed
turbulence may be viewed as the (quasi) thermodynamic equilibrium
of such a gas (or a solution of singularities in the background fluid).
We can define a temperature of turbulence, which is approximately
the mean kinetic energy of the translational motion of a singularity.
Also, we can estimate a chemical potential by evaluating $\overline{v^{2}}/2$.
We note that the density $\rho_{v}$ and the dimension $a$ of the
singularities remain undetermined; these parameters can be estimated
from experiment. 

Also, it is worth noting that we may consider the equations of the
fluid mechanics for this new gas of singularities, viewed as a continuous
medium, at a higher scale.

\section{Vorticial liquids}

\subsection{Vortex}

For an incompressible fluid with an isentropic flow we consider a
velocity field given by 
\begin{equation}
\boldsymbol{v}=\boldsymbol{\omega}\times gradf(r)\,\,\,,\label{80}
\end{equation}
where $\boldsymbol{\omega}$ is a vector which may depend only on
the time and the function \noun{$f(r)$ }is smooth everywhere, except,
possibly, at the origin $\boldsymbol{r}\neq0$, and vanishing rapidly
at infinity. The velocity $\boldsymbol{v}$ rotates about $\boldsymbol{\omega}$;
such a velocity field defines a vortex. 

We can ckeck that $div\boldsymbol{v}=0$ and 
\begin{equation}
\boldsymbol{v}=-curl\left[\boldsymbol{\omega}f(r)\right]\,\,\,,\label{81}
\end{equation}
 such that we can define a vector potential $\boldsymbol{A}=-\boldsymbol{\omega}f(r)$
($\boldsymbol{v}=curl\boldsymbol{A}$). We note that $div\boldsymbol{A}\neq0$
and the vorticity differs from $\Delta\boldsymbol{A}$, in general,
\begin{equation}
curl\boldsymbol{v}=\Delta\left[\boldsymbol{\omega}f(r)\right]-grad\,div\left[\boldsymbol{\omega}f(r)\right]\neq-\Delta\boldsymbol{A}\,\,.\label{82}
\end{equation}

If the velocity given by equation (\ref{81}) satisfies the second
equation (\ref{40}), we should have $\boldsymbol{\omega}\sim e^{-\nu\lambda t}$
and $\Delta f+\lambda f=0$, $i.e.$ $f\sim e^{\pm i\sqrt{\lambda}r}/r$,
where $\lambda$ is, in general, complex. In two dimensions $f$ is
a Bessel function. The Navier-Stokes equation gives the pressure. 

The most common example of a velocity field given by equation (\ref{80})
is the filamentary vortex (the \textquotedbl cyclon\textquotedbl ),
with $\boldsymbol{\omega}=const$, ($\lambda=0$), $f(r)=-\ln r$
and the two-dimensional position vector $\boldsymbol{r}$ perpendicular
to $\boldsymbol{\omega}$. In this case $\boldsymbol{v}=\boldsymbol{r}\times\boldsymbol{\omega}/r^{2}$,
$\boldsymbol{A}=\boldsymbol{\omega}\ln r$ and $curl\boldsymbol{v}=-2\pi\boldsymbol{\omega}\delta(\boldsymbol{r})$.
The velocity is singular at $\boldsymbol{r}=0$. For the filamentary
vortex the Navier-Stokes equation can be satisfied for $p=-\rho\omega^{2}/2r^{2}$,
\emph{i.e.} for an external potential $\varphi=p/\rho=-\omega^{2}/2r^{2}$.
This can be provided by the gravitational field for a fluid with a
free surface. 

For convenience, we give the expression of the various terms in the
Navier-Stokes equation for the vortex given by equation (\ref{80}), 

\begin{equation}
\begin{array}{c}
\frac{\partial\boldsymbol{v}}{\partial t}=\left(\dot{\boldsymbol{\omega}}\times\boldsymbol{r}\right)\frac{f^{'}}{r}\,\,,\\
\\
(\boldsymbol{v}grad)\boldsymbol{v}=\left[\boldsymbol{\omega}(\boldsymbol{\omega}\boldsymbol{r})-\omega^{2}\boldsymbol{r}\right]\frac{f^{'2}}{r^{2}}\,\,,\\
\\
\Delta\boldsymbol{v}=(\boldsymbol{\omega}\times\boldsymbol{r})\frac{(f^{''}+2f'/r)^{'}}{r}\,\,\,,
\end{array}\label{83}
\end{equation}
 where the primes denote the derivatives of the function $f;$ the
second equation (\ref{83}) is derived by using the identity
\begin{equation}
(\boldsymbol{v}grad)\boldsymbol{v}=-\boldsymbol{v}\times curl\boldsymbol{v}+grad(v^{2}/2)\,\,.\label{84}
\end{equation}
For a filamentary vortex the above expressions become 
\begin{equation}
\begin{array}{c}
(\boldsymbol{v}grad)\boldsymbol{v}=-\omega^{2}\boldsymbol{r}\frac{f^{'2}}{r^{2}}\,\,,\\
\\
\Delta\boldsymbol{v}=(\boldsymbol{\omega}\times\boldsymbol{r})\frac{(f^{''}+f'/r)^{'}}{r}\,\,.
\end{array}\label{85}
\end{equation}
In general, the vortex given by equation (\ref{80}) does not satisfy
the Navier-Stokes equation; it develops internal (Euler's) forces
which cannot be compensated; the vortex is unstable. Such vortices
are examples of singular velocities. 

\subsection{Vorticial liquid}

A set of vectors $\boldsymbol{\omega}_{i}$ placed at $\boldsymbol{r}_{i}$
form a vorticial liquid. The velocity field is
\begin{equation}
\boldsymbol{v}=\sum_{i}\boldsymbol{\omega}_{i}\times gradf_{i}(R_{i})\,\,\,,\label{86}
\end{equation}
 where $\boldsymbol{R}_{i}=\boldsymbol{r}-\boldsymbol{r}_{i}$. This
velocity field is a superpositon of independent vortices. Each $i$-th
vortex develops an internal force, while the inertial term generates
an interaction which brings an additional force; this additional force
depends on all the other $j$-th vortices, $j\neq i$. Therefore,
the Navier-Stokes equation is not satisfied by the velocity field
given by equation (\ref{86}), in the sense that there is no physical
external pressure to compensate the Euler force. 

We assume randomly fluctuating vectors $\boldsymbol{\omega}_{i}$,
as a distinctive feature of a fully developed vorticial turbulence.
It is likely that the fluid develops fluctuations as a reaction to
its uncompensated internal forces. Specifically, we assume $\overline{\boldsymbol{\omega}}_{i}=0$
and $\overline{\omega_{i}^{\alpha}\omega_{j}^{\beta}}=\frac{1}{3}\overline{\omega_{i}^{2}}\delta_{ij}\delta^{\alpha\beta}$,
where $\alpha,\,\beta=1,2,3$ are the cartesian labels of the components
of the vectors $\boldsymbol{\omega}_{i}$, and the overbar indicates
the average over time. Then, the average velocity is zero ($\overline{\boldsymbol{v}}=0$)
and we are left with the inertial term 
\begin{equation}
\overline{(\boldsymbol{v}grad)\boldsymbol{v}}=\overline{v_{j}\partial_{j}v_{i}}\,\,.\label{87}
\end{equation}
The calculation of this term is straightforward; we get 
\begin{equation}
\begin{array}{c}
\overline{(\boldsymbol{v}grad)\boldsymbol{v}}=\frac{1}{3}\sum_{i}\omega_{i}^{2}[\frac{1}{2}grad\left(gradf_{i}(R_{i})\right)^{2}-\\
\\
-gradf_{i}(R_{i})\cdot\Delta f_{i}(R_{i})]\,\,.
\end{array}\label{88}
\end{equation}
 The averaged Euler forces are 
\begin{equation}
\begin{array}{c}
-\overline{\boldsymbol{v}\times curl\boldsymbol{v}}=-\frac{1}{3}\sum_{i}\omega_{i}^{2}[\frac{1}{2}grad\left(gradf_{i}(R_{i})\right)^{2}+\\
\\
+gradf_{i}(R_{i})\cdot\Delta f_{i}(R_{i})]\,\,,\\
\\
\overline{grad(v^{2}/2)}=\frac{1}{3}\sum_{i}\omega_{i}^{2}grad\left(gradf_{i}(R_{i})\right)^{2}\,\,.
\end{array}\label{89}
\end{equation}
By using the spherical symmetry of the function $f_{i}(R_{i})$, these
expressions can be cast in the form 
\begin{equation}
\begin{array}{c}
\overline{(\boldsymbol{v}grad)\boldsymbol{v}}=-\frac{2}{3}\sum_{i}\omega_{i}^{2}f_{i}^{'2}\frac{\boldsymbol{R}_{i}}{R_{i}^{2}}\,\,,\\
\\
-\overline{\boldsymbol{v}\times curl\boldsymbol{v}}=-\frac{2}{3}\sum_{i}\omega_{i}^{2}f_{i}^{'}\left(f_{i}^{''}+\frac{1}{R_{i}}f_{i}^{'}\right)\frac{\boldsymbol{R}_{i}}{R_{i}}\,\,,\\
\\
\overline{grad(v^{2}/2)}=\frac{2}{3}\sum_{i}\omega_{i}^{2}f_{i}^{'}f_{i}^{''}\frac{\boldsymbol{R}_{i}}{R_{i}}\,\,.
\end{array}\label{90}
\end{equation}

We can see that even on average the Navier-Stokes equation is not
satisfied, in the sense discussed above. Even on average the vorticial
liquid develops internal forces which are not equilibrated; it is
unstable. We shall give specific examples of such an instability below. 

Now, let us assume that the vorticial liquid is sufficiently dense,
\emph{i.e.} if we can define a density $\rho_{v}$ of points $\boldsymbol{r}_{i}$;
further, we assume that the liquid is homogeneous and isotropic, \emph{i.e.}
this density is constant and the $\omega_{i}$ and the functions $f_{i}$
can be replaced in the above equations by uniform functions $\omega_{i}=\omega$
and $f_{i}=f$. We assume that this is another distinctive feature
of a fully developed turbulence. Then, we may transform the summation
over $i$ in equations (\ref{90}) into an integral, like in equation
(\ref{65}). By choosing the origin at $\boldsymbol{r}=\boldsymbol{r}_{i}$
for a fixed $\boldsymbol{r}_{i}$, and using the notation $\boldsymbol{R}=\boldsymbol{r}-\boldsymbol{r}_{i}$,
we get 
\begin{equation}
\overline{(\boldsymbol{v}grad)\boldsymbol{v}}=-\frac{2}{3}\rho_{v}\omega^{2}\int_{0}^{\infty}dR\cdot Rf^{'2}(R)\int do(\boldsymbol{R}/R)\,\,;\label{91}
\end{equation}
 the result of integration in this equation is zero, due to the integration
over the solid angle $o$, providing the radial integration is finite.
In this case the Navier-Stokes equation is satisfied trivially. 

Let us assume that the integral over $R$ is singular at $R=0$, as
another distinctive feature of a fully developed turbulence (the function
$f$ is assumed to decrease sufficiently rapid at infinity to have
a finite integral in this limit). The integration outside a small
region around the $i$-th point is zero, while the integration over
such a small region is indefinite. This singularity implies 
\begin{equation}
f(R)\sim(a/R)^{n}\,\,,\,\,n>0\label{02}
\end{equation}
 for $R\ll a$, where $a$ is a small characteristic distance (compare
with equation (\ref{66})). Therefore, we are left with a discrete
set of points $\boldsymbol{r}_{i}$, where the function $f(R_{i})$
and the velocity ($v\sim1/R_{i}^{n+1}$) are singular. We have now
a small region of dimension $a$, around each point $\boldsymbol{r}_{i}$,
which includes a mass of fluid, say, $M$, where the inertial term
given by equation (\ref{91}) is not defined. Since the positions
$\boldsymbol{r}_{i}$ may change in time, we are left with a classical
gas of particle-like vortices (or a solution of vortices in the background
fluid), as discussed above. Of course, we may have also a mixture
of vorticial gases, each characterized by a dimension $a$ and a mass
$M$.

The equation of energy conservation (equation (\ref{67})) 
\begin{equation}
\frac{\partial}{\partial t}\left(\frac{1}{2}v^{2}\right)+\boldsymbol{v}\cdot(\boldsymbol{v}grad)\boldsymbol{v}=\nu\boldsymbol{v}\Delta\boldsymbol{v}\,\,,\label{93}
\end{equation}
 averaged over the fluctuating vectors $\boldsymbol{\omega}_{i}$,
is reduced to the viscosity term 
\begin{equation}
\nu\overline{\boldsymbol{v}\Delta\boldsymbol{v}}=\frac{2\nu}{3}\sum_{i}\omega_{i}^{2}f_{i}^{'}\left(f_{i}^{''}+2f_{i}^{'}/R_{i}\right)^{'}\,\,.\label{94}
\end{equation}
This term should be computed outside the regions with dimension $a$,
where the motion is defined. The non-vanishing value of this term
indicates an energy loss, which should be compensated from the outside.
The vorticial gas is in quasi equilibrium. 

The above considerations are valid for spherical-symmetric functions
$f_{i}(R_{i})$; if the vortices have a lower (internal) symmetry,
the unit vector $\boldsymbol{R}/R$ in equation (\ref{91}) is replaced
by functions which do not have a spherical symmetry, and the inertial
term is not vanishing, in general. The vortices are unstable, and,
likely, they could tend to acquire a spherical symmetry, which ensures
a (quasi)-equilibrium. 

\subsection{Filamentary liquid}

Let us consider a set of rectilinear, parallel filaments, directed
along the $z$-axis, placed at positions $\boldsymbol{r}_{i}$ in
the $(x,y)$-plane, with vorticities $\boldsymbol{\omega}_{i}$. This
is a two-dimensional vorticial liquid of \textquotedbl cyclons\textquotedbl .
The velocity field is given by 
\begin{equation}
\boldsymbol{v}=-\sum_{i}\boldsymbol{\omega}_{i}\times grad\ln R_{i}=\sum_{i}\frac{\boldsymbol{R}_{i}\times\boldsymbol{\omega}_{i}}{R_{i}^{2}}\,\,\,,\label{95}
\end{equation}
 where $\boldsymbol{R}_{i}=\boldsymbol{r}-\boldsymbol{r}_{i}$. 

Equation (\ref{95}) shows that the velocity is derived from a vector
potential $\boldsymbol{A}$, through $\boldsymbol{v}=curl\boldsymbol{A}$.
By taking the $curl$ in this equation, we get 
\begin{equation}
\Delta\boldsymbol{A}=-2\boldsymbol{\omega}\label{96}
\end{equation}
(providing $div\boldsymbol{A}=0$ and $div\boldsymbol{\omega}=0$),
where we introduce the notation $curl\boldsymbol{v}=2\boldsymbol{\omega}$;
$\boldsymbol{\omega}$ is called vorticity. The vorticity distribution
corresponding to equation (\ref{95}) 
\begin{equation}
\boldsymbol{\omega}(\boldsymbol{r})=\frac{1}{2}curl\boldsymbol{v}=-\pi\sum_{i}\boldsymbol{\omega}_{i}\delta(\boldsymbol{R}_{i})\,\,\,,\label{97}
\end{equation}
gives the vector potential 
\begin{equation}
\boldsymbol{A}=\sum_{i}\boldsymbol{\omega}_{i}\ln R_{i}\,\,.\label{98}
\end{equation}
According to equation (\ref{84}), the inertial term has the components
\begin{equation}
\boldsymbol{f}=-\boldsymbol{v}\times curl\boldsymbol{v}=-2\pi\sum_{i\neq j}\omega_{i}\omega_{j}grad_{i}\ln R_{ij}\cdot\delta(\boldsymbol{R}_{i})\,\,\,,\label{99}
\end{equation}
 where $\boldsymbol{R}_{ij}=\boldsymbol{r}_{i}-\boldsymbol{r}_{j}$,
and $grade$, where 
\begin{equation}
e=\frac{1}{2}v^{2}=\sum_{i\neq j}\omega_{i}\omega_{j}\frac{\boldsymbol{R}_{i}\boldsymbol{R}_{j}}{2R_{i}^{2}R_{j}^{2}}\label{100}
\end{equation}
 is a density of kinetic energy (per unit mass). Apart from the force
$\boldsymbol{f}$, which acts at the positions of the vortices, there
exist internal forces given by $grade$, which make the liquid unstable.
The motion and the statistics of parallel, rectilinear filaments have
been extensively investigated,\cite{key-24}-\cite{key-31} the instability
being associated with a negative temperature in an attempt of a statistical
theory.\cite{key-25,key-27}

The total force given by equation (\ref{99}) 
\begin{equation}
\boldsymbol{F}=\int d\boldsymbol{r}\boldsymbol{f}=-2\pi\sum_{i\neq j}\omega_{i}\omega_{j}grad_{i}\ln R_{ij}\label{101}
\end{equation}
 is zero. We can see that a force 
\begin{equation}
\boldsymbol{f}_{ij}=-\boldsymbol{f}_{ji}=-2\pi\omega_{i}\omega_{j}grad_{i}\ln R_{ij}\label{102}
\end{equation}
 acts between any pair $(ij)$ of vortices. This force derives from
a potential 
\begin{equation}
U_{ij}=2\pi\omega_{i}\omega_{j}\ln R_{ij}\,\,\,,\label{103}
\end{equation}
 such that 
\begin{equation}
\boldsymbol{F}=-\sum_{i\neq j}grad_{i}U_{ij}\label{104}
\end{equation}
 The density of kinetic energy $e$ can be written as 
\begin{equation}
e=\frac{1}{2}v^{2}=\frac{1}{2}\boldsymbol{v}curl\boldsymbol{A}=-\frac{1}{2}div(\boldsymbol{v}\times\boldsymbol{A})+\boldsymbol{A}\boldsymbol{\omega}\,\,.\label{105}
\end{equation}
 The first term in equation (\ref{105}) is singular at $\boldsymbol{r}=\boldsymbol{r}_{i}$;
we integrate this term over the whole space, transform it into surface
integrals, both at infinity and over small circles around each filament,
and neglect their contributions. The result of such integrations is
a self-energy (or a self-force), which may be left aside. We call
this procedure a \textquotedbl renormalization\textquotedbl .\cite{key-32}
By doing so, we are left with a total kinetic energy
\begin{equation}
\begin{array}{c}
E=\int d\boldsymbol{r}e=\int d\boldsymbol{r}\boldsymbol{A}\boldsymbol{\omega}=-\pi\sum_{i\neq j}\omega_{i}\omega_{j}\ln R_{ij}=\\
\\
=-\frac{1}{2}\sum_{i\neq j}U_{ij}=-U\,\,\,,
\end{array}\label{106}
\end{equation}
 where $U$ is the total potential energy. We can see that the total
energy is conserved, \emph{i.e.} $E+U=const$. Also, the total force
$\boldsymbol{F}=0$, such that the total momentum is conserved. The
total angular momentum is $-2\int d\boldsymbol{r}\boldsymbol{A}$;
it is proportional to $\sum_{i}\boldsymbol{\omega}_{i}$. The total
torque 
\begin{equation}
2\pi\sum_{i\neq j}\omega_{i}\omega_{j}\frac{\boldsymbol{r}_{i}\times\boldsymbol{r}_{j}}{R_{ij}^{2}}\label{107}
\end{equation}
 is zero. By this \textquotedbl renormalization\textquotedbl{} procedure
the points $\boldsymbol{r_{i}}$ are completely decoupled from the
fluid, and they may have their own motion. 

The average over fluctuating vorticities can be computed straightforwardly,
by using $\overline{\boldsymbol{\omega}_{i}^{2}}=\frac{1}{2}\omega_{i}^{2}$;
it is given by 
\begin{equation}
\overline{(\boldsymbol{v}grad)\boldsymbol{v}}=-\sum_{i}\omega_{i}^{2}\frac{\boldsymbol{R}_{i}}{2R_{i}^{4}}\,\,.\label{108}
\end{equation}
 We can see that for a sufficiently dense, homogeneous and isotropic
liquid we get a gas of (singular) vortices, as discussed above. 

\subsection{Coulombian liquid}

For $f_{i}(R_{i})=-1/R_{i}$ in equation (\ref{86}) we get a coulombian
vorticial liquid with the velocity field 
\begin{equation}
\boldsymbol{v}=-\sum_{i}\boldsymbol{\omega}_{i}\times grad(1/R_{i})=\sum_{i}\frac{\boldsymbol{\omega}_{i}\times\boldsymbol{R}_{i}}{R_{i}^{3}}\label{109}
\end{equation}

and the vector potential 
\begin{equation}
\boldsymbol{A}=\sum_{i}\frac{\boldsymbol{\omega}_{i}}{R_{i}}\,\,.\label{110}
\end{equation}
The equation $\Delta\boldsymbol{A}=-2\boldsymbol{\omega}$ is satisfied
for
\begin{equation}
\boldsymbol{\omega}(\boldsymbol{r})=2\pi\sum_{i}\boldsymbol{\omega}_{i}\delta(\boldsymbol{R}_{i})\,\,\,,\label{111}
\end{equation}
 but this vorticity differs from 
\begin{equation}
\frac{1}{2}curl\boldsymbol{v}=2\pi\sum_{i}\boldsymbol{\omega}_{i}\delta(\boldsymbol{R}_{i})-\frac{1}{2}grad\,div\sum_{i}(\boldsymbol{\omega}_{i}/R_{i})\,\,.\label{112}
\end{equation}

By applying the \textquotedbl renormalization\textquotedbl{} procedure
we get the force
\begin{equation}
\boldsymbol{F}=\int d\boldsymbol{r}\boldsymbol{f}=-\int dr\boldsymbol{v}\times curl\boldsymbol{v}=4\pi\sum_{i\neq j}grad_{i}\frac{\boldsymbol{\omega}_{i}\boldsymbol{\omega}_{j}}{R_{ij}}\label{113}
\end{equation}
and the energy
\begin{equation}
E=\int d\boldsymbol{r}e=\frac{1}{2}\int drv^{2}=2\pi\sum_{i\neq j}\frac{\boldsymbol{\omega}_{i}\boldsymbol{\omega}_{j}}{R_{ij}}\,\,\,,\label{114}
\end{equation}
 where, in both cases $\boldsymbol{v}=curl\boldsymbol{A}$ is used.
The potential from equation (\ref{103}) is now $U_{ij}=-4\pi\frac{\boldsymbol{\omega}_{i}\boldsymbol{\omega}_{j}}{R_{ij}}$.
The total energy is conserved, the total force is zero, the total
angular momentum is proportional to $\sum_{i}\boldsymbol{\omega}_{i}$
and the total torque (zero) is 
\begin{equation}
4\pi\sum_{i\neq j}\boldsymbol{\omega}_{i}\boldsymbol{\omega}_{j}\frac{\boldsymbol{r}_{i}\times\boldsymbol{r}_{j}}{R_{ij}^{3}}\,\,.\label{115}
\end{equation}

The average of the inertial term over fluctuating vortices is obtained
from equation (\ref{90})
\begin{equation}
\overline{(\boldsymbol{v}grad)\boldsymbol{v}}=-\frac{2}{3}\sum_{i}\omega_{i}^{2}\frac{\boldsymbol{R}_{i}}{R_{i}^{6}}\,\,\,,\label{116}
\end{equation}
such that we may get a gas of singular vortices.

\subsection{Dipolar liquid}

A dipolar liquid is defined by the vorticity
\begin{equation}
\boldsymbol{\omega}=-2\pi\sum_{i}\boldsymbol{m}_{i}\times grad\delta(\boldsymbol{r}-\boldsymbol{r}_{i})\,\,\,,\label{117}
\end{equation}
 where the vectors $\boldsymbol{m}_{i}$ may depend on the time, at
most. We get the vector potential 
\begin{equation}
\boldsymbol{A}=\sum_{i}\frac{\boldsymbol{m}_{i}\times\boldsymbol{R}_{i}}{R_{i}^{3}}=-\sum_{i}\boldsymbol{m}_{i}\times grad(1/R_{i})\label{118}
\end{equation}
 and the velocity field 
\begin{equation}
\begin{array}{c}
\boldsymbol{v}(\boldsymbol{r})=\sum_{i}\left[-\boldsymbol{m}_{i}/R_{i}^{3}+3\boldsymbol{R}_{i}(\boldsymbol{m}_{i}\boldsymbol{R}_{i})/R_{i}^{5}\right]=\\
\\
=\sum_{i}grad\left[\boldsymbol{m}_{i}grad(1/R_{i})\right]\,\,.
\end{array}\label{119}
\end{equation}
We recognize in these equations magnetic (dipole) moments $\boldsymbol{m}_{i}$,
a dipolar vector potential $\boldsymbol{A}$ and a magnetic field
$\boldsymbol{v}$. 

The inertial term has the components 
\begin{equation}
\begin{array}{c}
\boldsymbol{f}=-\boldsymbol{v}\times curl\boldsymbol{v}=2\boldsymbol{\omega}\times\boldsymbol{v}=\\
\\
=4\pi\sum_{i\neq j}grad\left[\boldsymbol{m}_{j}grad(1/R_{j})\right]\times\left[\boldsymbol{m}_{i}\times grad\delta(\boldsymbol{R}_{i})\right]
\end{array}\label{120}
\end{equation}
and $grade$, where 
\begin{equation}
\begin{array}{c}
e=\frac{1}{2}v^{2}=\frac{1}{2}\sum_{i\neq j}[\frac{\boldsymbol{m}_{i}\boldsymbol{m}_{j}}{R_{i}^{3}R_{j}^{3}}-\frac{3(\boldsymbol{m}_{i}\boldsymbol{R}_{j})(\boldsymbol{m}_{j}\boldsymbol{R}_{j})}{R_{i}^{3}R_{j}^{5}}-\\
\\
-\frac{3(\boldsymbol{m}_{j}\boldsymbol{R}_{i})(\boldsymbol{m}_{i}\boldsymbol{R}_{i})}{R_{i}^{5}R_{j}^{3}}+\\
\\
+\frac{9(\boldsymbol{R}_{i}\boldsymbol{R}_{j})(\boldsymbol{m}_{i}\boldsymbol{R}_{i})(\boldsymbol{m}_{j}\boldsymbol{R}_{j})}{R_{i}^{5}R_{j}^{5}}]\,\,.
\end{array}\label{121}
\end{equation}
 We can see that the dipolar liquid is unstable. 

The total force and the total kinetic energy are 
\begin{equation}
\begin{array}{c}
\boldsymbol{F}=\int d\boldsymbol{r}\boldsymbol{f}=2\int d\boldsymbol{r}\boldsymbol{\omega}\times\boldsymbol{v}=-\sum_{i\neq j}grad_{i}U_{ij}\,\,\,,\\
\\
E=\int d\boldsymbol{r}e=\int d\boldsymbol{r}\boldsymbol{\omega}\boldsymbol{A}=-\frac{1}{2}\sum_{i\neq j}U_{ij}\,\,\,,
\end{array}\label{122}
\end{equation}
where 
\begin{equation}
U_{ij}=-4\pi\boldsymbol{m}_{i}grad_{i}\left[\boldsymbol{m}_{j}grad_{i}(1/R_{ij})\right]\,\,.\label{123}
\end{equation}
By this \textquotedbl renormalization\textquotedbl{} procedure, the
liquid is reduced to a set of interacting particle-like vortices.
We can see that the energy is conserved, the total force is zero,
the total angular momentum and the total torque are zero.

The time average over vorticities in equation (\ref{121}) leads to
\begin{equation}
\overline{e}=\frac{1}{2}\sum_{i}\left[\frac{m_{i}^{2}}{R_{i}^{6}}+\frac{3\overline{(\boldsymbol{m}_{i}\boldsymbol{R}_{i})^{2}}}{R_{i}^{8}}\right]=\sum_{i}\frac{m_{i}^{2}}{R_{i}^{6}}\,\,\,,\label{124}
\end{equation}
which gives a force 
\begin{equation}
grad\overline{e}=-6\sum_{i}\frac{m_{i}^{2}\boldsymbol{R}_{i}}{R_{i}^{8}}\,\,.\label{125}
\end{equation}

For a dense, homogeneous and isotropic liquid the force given by equation
(\ref{125}) is zero (for $\boldsymbol{R}_{i}\neq0$); also, the force
$\boldsymbol{f}$ is zero for $\boldsymbol{R}_{i}\neq0$, and the
Navier-Stokes equation is satisfied trivially (on average). We note
that $\Delta\boldsymbol{v}$ is zero for $\boldsymbol{R}_{i}\neq0$
(equation (\ref{119})), such that the viscosity contribution is zero.
We are left with a set of positions $\boldsymbol{r}_{i}$, each surrounded
by a small region, where the motion is not defined. According to the
above discussion, such a structure may be viewed as a (quasi) ideal
classical gas of vortices (or a solution of vortices in the background
fluid).

\section{Concluding remarks}

In fairly general conditions we have given in this paper an explicit
(smooth) solution for the potential flow. We have shown that, rigorously
speaking, the equations of the fluid mechanics have not rotational
solutions. However, usually we may neglect the variations of the density
and the temperature, such that, in these conditions, the Navier-Stokes
equation may exhibit (approximate) vorticial solutions, governed by
the viscosity. We give arguments that the vortices are unstable. On
the other hand, for large variations of the velocity over small distances,
the fluid velocity exhibits turbulent, highly fluctuating instabilities,
controlled by viscosity. Such a fully developed turbulence occurs
as a consequence of the insufficiency of the viscosity to dissipate
heat. We represent the fully developed turbulence as a superposition
of highly fluctuating velocities, associated to a discrete distrbution
of turbulence centres, and are interested in the temporal average
of this velocity field. A regular mean flow may be added. It is shown
that the Navier-Stokes equation is not satisfied on average. However,
for a homogeneous and isotropic distribution of (non-singular) turbulence
centres (as another distinctive feature of a fully developed turbulence),
the temporal average of both the fluctuating velocity and the inertial
term is zero, such that the Navier-Stokes equation is satisfied trivially.
If the velocity is singular at the turbulence centres we are left
with a quasi ideal classical gas of singularities (or a solution of
singularities in the background fluid), in thermal quasi equilibrium,
as an example of emergent dynamics. The Navier-Stokes equation for
this fluid of singularities is reduced to Newton's law of motion (with
a small friction). At a higher scale, equations of fluid mechanics
can be considerd for this gas, as a continuous medium. We have illustrated
all the above considerations with three examples of (singular) vorticial
liquids.

\textbf{Acknowledgements}

The author is indebted to the members of the Department of Theoretical
Physics, the Institute of Physics and Nuclear Engineering, the Institute
of Atomic Physics, Magurele, for many enlightening discussions. A
helpful analysis by dr. F. Buzatu is particularly acknowledged. This
work was carried out within the Program Nucleu, funded by the Romanian
Ministry of Research, Innovation and Digitization, projects no. PN23210101/2023.

\textbf{Conflict of interests:} The author declare no conflict of
interest.

\end{document}